\documentclass[sigconf]{acmart}   
\renewcommand\footnotetextcopyrightpermission[1]{}
\acmConference[ASP-DAC '27]{the 32nd Asia and South Pacific Design Automation
  Conference}{January 25--28, 2027}{Tokyo, Japan}
\acmYear{2027}
\copyrightyear{2026}

\usepackage{graphicx}
\usepackage{subcaption}
\usepackage{booktabs}
\usepackage{xcolor}
\usepackage{amsmath}

\newif\ifshowtbd\showtbdtrue        

\begin{document}

\title{Compiling Semi-Ring Dynamic Programming to Tier-Aware 3D-DRAM Processing-in-Memory}

\author{Mahbod Afarin}
\affiliation{%
  \institution{University of California San Diego}
  \city{La Jolla}
  \state{California}
  \country{USA}}
\email{mafarin@ucsd.edu}

\author{Tsung-Han Lu}
\affiliation{%
  \institution{University of California San Diego}
  \city{La Jolla}
  \state{California}
  \country{USA}}
\email{tsl012@ucsd.edu}

\author{Tajana Rosing}
\affiliation{%
  \institution{University of California San Diego}
  \city{La Jolla}
  \state{California}
  \country{USA}}
\email{tajana@ucsd.edu}

\begin{abstract}
Processing-in-Memory (PIM) on monolithic 3D (M3D) DRAM is a promising answer to
the memory wall for data-intensive dynamic programming (DP), yet extracting its
performance today demands hand-written kernels: the programmer must pick a tile
size, place data across non-uniform-latency memory tiers, partition work between
heterogeneous processing units, and insert the right broadcasts and queues by
hand. We present \emph{GenMLIR}, an MLIR compiler that automates this mapping for
tier-aware 3D PIM. GenMLIR encodes the semi-ring generalized grid update, the
algebraic form shared by all-pairs shortest path (APSP) and sequence alignment,
as a first-class IR abstraction, and lowers it through four GenDRAM-aware pass
groups that derive blocked tiling, tier-aware placement, tile-to-PU assignment,
and explicit communication. We also characterize precisely which DP recurrences
the abstraction admits and which it does not. On the GenDRAM architecture,
GenMLIR-generated code runs up to 5.8$\times$ (APSP) and 16.7$\times$ (alignment)
faster than a lowering without compiler support, and 1.2--1.4$\times$ faster than
a standard affine-tiling PIM compiler we also implement, reaching 90--100\% of an
achievable-performance bound. It expresses these workloads in $4$--$9$ lines
instead of ${\sim}300$--$500$ and compiles in a negligible fraction of runtime.
\end{abstract}

\keywords{Processing-in-Memory, MLIR, Monolithic 3D DRAM, Dynamic Programming,
Compiler, Tier-Aware Data Placement}

\maketitle
\pagestyle{plain}

\section{Introduction}
\label{sec:introduction}

Monolithic 3D DRAM bonds a logic die directly beneath a stack of DRAM layers,
exposing tens of TB/s of internal bandwidth to compute units placed next to the
banks~\cite{pan2025stratum}. This makes it an attractive substrate for dynamic
programming (DP) workloads such as all-pairs shortest path (APSP) and genomic
sequence alignment, whose runtime is dominated by data movement rather than
arithmetic~\cite{li2018minimap2,ahn2015pim,dai2018graphh}. Recent PIM
accelerators exploit exactly this~\cite{xu2023rapidx,lu2025gendram}.

The hardware, however, has outrun the software. A modern 3D PIM part such as
GenDRAM~\cite{lu2025gendram} presents the programmer with a stack whose
row-access latency varies by $1.6\times$ from the bottom tier to the top, two functionally
distinct classes of processing unit, an inter-PU ring that must be driven in
either broadcast or pipelined mode, and a bank-group geometry that dictates how
tiles may be laid out. Achieving good performance means choosing a tile size
against an on-chip SRAM budget, pinning latency-critical tables to fast tiers
while interleaving bandwidth-bound tensors across channels, assigning tiles to
PUs so that neighbors do not collide in the same bank, and inserting broadcasts,
barriers, and queues that respect the recurrence. Today all of this is done by
hand, once per kernel. The resulting code is unportable and, more importantly,
it is not reusable: a change of recurrence means starting over.

Our starting point is an observation made by the GenDRAM
authors~\cite{lu2025gendram} and, before them, by the GraphBLAS
community~\cite{kepner2016graphblas}: APSP and sequence alignment are the same
computation over different semi-rings. Both repeatedly update a cell of a grid
by combining ($\otimes$) two neighbors and accumulating ($\oplus$) the result.
GenDRAM used this isomorphism to justify a \emph{single hardware substrate}. We
argue it also supports a single \emph{compilation} strategy---and that this is
where the abstraction pays off most, because the tiling, dependence, placement,
and communication decisions listed above are all independent of which semi-ring
is instantiated. What is workload-specific is only the execution mode.

This paper turns that argument into a compiler. \textbf{GenMLIR} is an MLIR
stack that takes a semi-ring DP recurrence plus a small configuration record and
emits a GenDRAM command stream. Its contributions are:

\begin{itemize}\itemsep2pt
  \item \textbf{An IR for tier-aware DP.} We encode the generalized grid update
  as an MLIR dialect and pair it with a device-level dialect that exposes
  tiers, bank-groups, PU classes, and communication primitives, so that
  placement and scheduling become IR rewrites rather than hand edits
  (\S\ref{sec:compiler}).
  \item \textbf{A characterization of the abstraction's reach.} We identify the
  algebraic conditions a recurrence must satisfy to be admitted, enumerate seven
  DP problems that qualify, and state explicitly which ones do not and why
  (\S\ref{sec:generality}). This scopes a claim that was previously left open.
  \item \textbf{Tier-aware and heterogeneity-aware passes.} We give a capacity
  model that derives the tile size directly from the PU SRAM budget, a tier
  assignment driven by access-latency sensitivity, and a mode-selection pass
  that chooses between broadcast and producer--consumer lowering
  (\S\ref{sec:midend}). We separate these from the standard transformations we
  reuse.
  \item \textbf{An evaluation against strong baselines.} We compare against
  an achievable bound and a standard affine-tiling PIM
  pass, not only against a lowering without compiler support,
  and add a memory-traffic analysis and compile-time
  and lines-of-code measurements (\S\ref{sec:results}).
\end{itemize}


\begin{figure*}[t]
    \centering
    \includegraphics[width=0.87\linewidth]{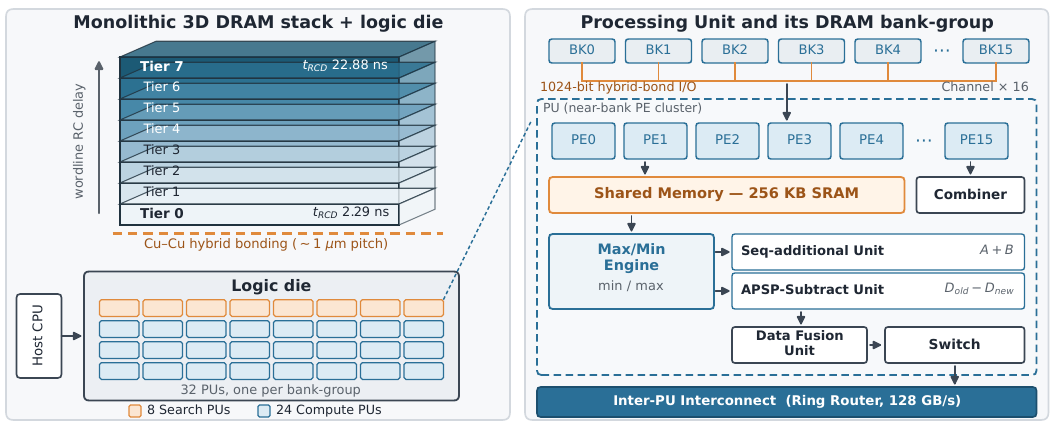}
    \vspace{-0.1in}
    \caption{The GenDRAM target. \textbf{Left:} an M3D DRAM stack of eight
    latency tiers ($t_{\mathrm{RCD}}$ 2.29--22.88\,ns; wordline RC delay grows
    with distance from the logic die), bonded at $\sim$1\,$\mu$m pitch to a logic
    die holding 32 PUs---one per bank-group, since 16 channels $\times$ 2
    bank-groups per channel $=32$. \textbf{Right:} a PU (dashed) is a near-bank
    PE cluster of 16 lightweight PEs, a 256\,KB shared scratchpad and combiner, a
    multiplier-less Max/Min engine with its Seq-additional and APSP-subtract
    sub-units, and an interconnect interface (data fusion unit and switch). The
    DRAM banks, the bank/channel bus, and the inter-PU ring lie outside the PU.
    Redrawn after Fig.~8 of~\cite{lu2025gendram}.}
    \label{fig:gendram_overview}
    \vspace{-0.12in}
\end{figure*}

\section{Background and Related Work}
\label{sec:background}

\subsection{Dynamic Programming as a Semi-Ring}
\label{sec:semiring}

A semi-ring is a set equipped with an associative combination operator $\otimes$
and an associative, commutative accumulation operator $\oplus$ over which
$\otimes$ distributes. A large family of DP algorithms updates a matrix $D$ by

\begin{equation}
\label{eq:semiring}
D[i,j] \;\leftarrow\; D[i,j] \;\oplus\; \bigl( D[i,k] \otimes D[k,j] \bigr),
\end{equation}

instantiating $(\oplus,\otimes)$ per domain. Floyd--Warshall APSP relaxes edges
over the $(\min,+)$ semi-ring; Smith--Waterman alignment maximizes a score over
$(\max,+)$. The two are structurally identical and differ only in the operator
pair and in how the resulting dependence graph is traversed---globally, in
pivot-broadcast super-steps, versus locally, along a diagonal wavefront. This
algebraic unification is long-established in the graph-analytics
literature~\cite{kepner2016graphblas} and was applied to 3D PIM hardware design
by GenDRAM~\cite{lu2025gendram}; Eq.~\eqref{eq:semiring} is reproduced from the
latter. Our contribution is not the abstraction but its use as a compiler IR,
together with a precise statement of its limits (\S\ref{sec:generality}).

\subsection{Target Architecture: GenDRAM}
\label{sec:gendram}

\vspace{0.04 in}
\noindent\textbf{Relationship to prior work.}
GenMLIR targets GenDRAM~\cite{lu2025gendram}, a recently 
proposed M3D DRAM PIM accelerator. GenDRAM
contributes \emph{all} of the hardware in this subsection: the tiered
stack, the heterogeneous PU organization, the modulo tile-to-PU mapping rule of
Eq.~\eqref{eq:mapping}, the decision to pin the PTR/CAL tables to Tier~0, the
two execution modes, and the cycle-accurate simulator used in
\S\ref{sec:results}. Fig.~\ref{fig:gendram_overview} is redrawn after Fig.~8
of~\cite{lu2025gendram}. What
GenDRAM does \emph{not} provide, and what this paper supplies, is a way to reach
any of it from a program rather than from a hand-written mapping.

\vspace{0.04 in}
\noindent\textbf{Structure.}
The logic die carries 32 PUs in a 1:1 coupling with the DRAM bank-groups, so an
\texttt{ACTIVATE} delivers a full row into the owning PU's scratchpad over the
hybrid-bond interface. Eight PUs are \emph{Search PUs}, built for the irregular,
latency-critical PTR$\rightarrow$CAL lookups of genomic seeding; the remaining
24 are \emph{Compute PUs}, each a 16-PE cluster whose reconfigurable
adder/comparator implements $(\min,+)$ or $(\max,+)$ directly. Three properties
drive every compiler decision that follows:

\emph{(i) Tiered latency.} The staircase wordline routing of a 1024-layer stack
makes $t_{\mathrm{RCD}}$ rise ${\sim}10\times$ (2.29 to 22.88\,ns), widening
full row-access time $t_{\mathrm{RC}}$ by $1.6\times$ (35 to 55\,ns). Placement is therefore not
latency-neutral.

\emph{(ii) Bank-group geometry.} A tile's owner PU is fixed by
\begin{equation}
\label{eq:mapping}
\mathrm{PU}(i,j) = (i \times M + j) \bmod (C \times G),
\end{equation}
with $M$ tiles per row, $C$ channels and $G$ bank-groups per channel, so
logically adjacent tiles land on distinct PUs and wavefront neighbors do not
contend.

\emph{(iii) Two execution modes.} APSP runs homogeneously: a pivot PU updates
$A_{kk}$, broadcasts it over the ring, and all PUs then update row/column and
internal tiles in lockstep. Genomics runs as a heterogeneous pipeline: Search
PUs stream candidate locations into queues consumed by Compute PUs.

\subsection{MLIR}
MLIR~\cite{lattner2021mlir} supports coexisting abstraction levels in one
pipeline via user-defined dialects that interoperate with existing passes.
Halide~\cite{ragankelley2013halide} and TVM~\cite{chen2018tvm} showed the value
of separating algorithm from schedule; MLIR makes that separation extensible. It
fits here because the object being lowered, a grid update, must successively
acquire a tile structure, an owner PU, a memory tier, and a communication
schedule, each naturally a distinct IR level rather than a compiler-internal
side table.

\subsection{Related Work}
\label{sec:related}

\vspace{0.04 in}
\noindent\textbf{Synthesis of DP hardware.}
ROCCC generates systolic arrays for wavefront algorithms from C~\cite{Buyukkurt08};
DWARV derives pipelined DP datapaths by recursive variable expansion~\cite{Nawaz09};
Cinnabar synthesizes DP energy controllers~\cite{Cinnabar}; and DP-HLS supplies a
C++ template library over Vitis HLS for bioinformatics kernels~\cite{DP-HLS}. All
emit a frozen hardware instance per recurrence, so any algorithmic change costs a
full synthesis cycle. GenMLIR instead targets fixed, programmable PIM hardware.

\vspace{0.04 in}
\noindent\textbf{Mapping onto programmable accelerators.}
GenDP couples a generic DP accelerator with DPMap, which lowers data-flow graphs
to control instructions \cite{GenDP}; GMX adds DP-oriented ISA extensions to
RISC-V \cite{GMX}; Darwin-WGA exposes a co-processor through
OpenCL \cite{DarwinWGA,DallyDSA}; GenASM accelerates approximate string
matching \cite{cali2020genasm}. All assume a flat, uniform-latency memory. None
must decide which of eight tiers a table belongs in, or how to split a kernel
across two PU classes, the two decisions that dominate performance on M3D PIM,
and the reason a new pass design is needed rather than a port. Likewise,
HPVM-HDC compiles across CPUs, GPUs and accelerators from a heterogeneous
IR~\cite{HPVM-HDC}; we adopt its philosophy, but to our knowledge no existing
MLIR pipeline models non-uniform DRAM tier latency or heterogeneous near-bank PU
classes as first-class IR concepts.

\section{The GenMLIR Compiler}
\label{sec:compiler}

\begin{figure*}[t]
    \centering
    \includegraphics[width=0.89\linewidth]{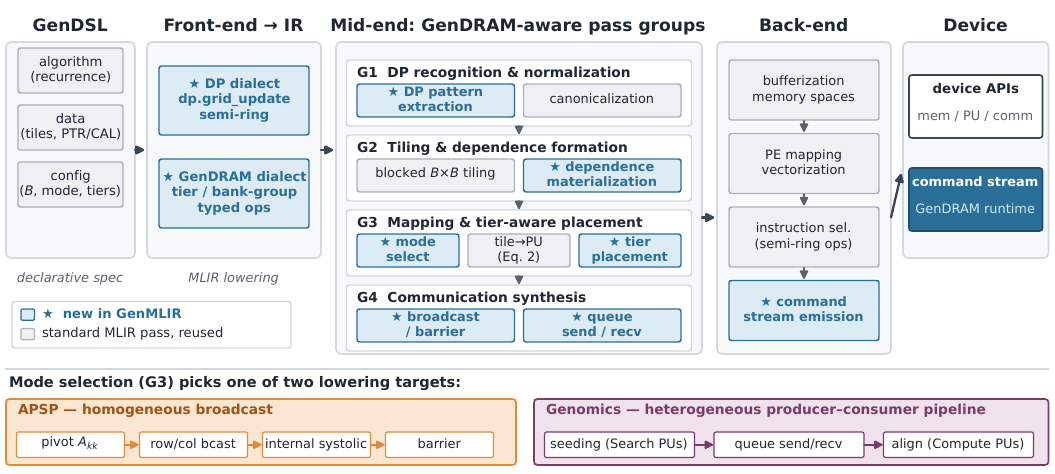}
    \vspace{-0.1in}
    \caption{The GenMLIR compilation pipeline. A GenDSL specification is lowered
    into a DP-level dialect and a GenDRAM device dialect, transformed by four
    mid-end pass groups, and emitted as a command stream. Passes marked
    $\bigstar$ are new to this work; unmarked passes are general-purpose MLIR
    transformations GenMLIR builds on. The bottom band shows the two
    execution modes that mode selection (G3) chooses between.}
    \label{fig:passes}
    \vspace{-0.12in}
\end{figure*}

GenMLIR lowers a DP specification to a GenDRAM command stream in three phases:
a front-end that captures algorithm, data, and configuration; a mid-end of four
GenDRAM-aware pass groups; and a back-end that emits device API calls and
command streams. Fig.~\ref{fig:passes} shows the full pipeline.

\subsection{What Is New, and What Is Reused}
\label{sec:novelty}

Because a compiler is assembled largely from known transformations, we state
the boundary explicitly rather than leave it to inference; it is also drawn in
Fig.~\ref{fig:passes}.

\emph{General-purpose MLIR passes:} loop canonicalization, blocked
tiling of an affine nest, bufferization and memory-space assignment, SIMD
lowering onto PE lanes, and instruction selection. These form the lowering
backbone that the passes below extend with DP and tier-specific reasoning.

\emph{New to this work:} (1) \emph{DP pattern extraction}, which recovers a
semi-ring \texttt{dp.grid\_update} from a scalar add/compare/select nest, so
that downstream passes reason about an algebraic operator rather than a fixed
opcode sequence; (2) \emph{dependence materialization}, which synthesizes either
the three-phase blocked Floyd--Warshall dependence structure or a diagonal
wavefront order from the same \texttt{dp.grid\_update}, selected by operator
signature; (3) \emph{tier-aware placement}, which ranks memory objects by
latency sensitivity and assigns them to physical tiers under a capacity
constraint (\S\ref{sec:midend}); (4) \emph{mode selection}, which commits the
module to broadcast or producer--consumer lowering; and (5) \emph{communication
synthesis}, which turns ownership plus dependence edges into ring broadcasts,
barriers, or queue send/recv pairs. Items (1)--(2) are what make the semi-ring
abstraction operational; (3)--(5) are what make it \emph{tier-aware}, and have
no analogue in flat-memory accelerator compilers.

\subsection{Front-End: GenDSL}
GenDSL is a small declarative language with three sections. \emph{Algorithm}
names the semi-ring $(\oplus,\otimes)$ and the recurrence's index pattern.
\emph{Data} declares the matrices, the PTR/CAL lookup tables, and the
producer--consumer streams between seeding and alignment. \emph{Configuration}
exposes the knobs a user may wish to override: tile size $B$, execution mode, the
Search/Compute PU split, tier hints, and datapath precision. These all default to
compiler-derived values, so a correct program need not mention them. GenDSL
lowers to two dialects: a \emph{DP dialect} that makes the recurrence and the
intended traversal explicit, and a \emph{GenDRAM dialect} whose types carry a
tier, a channel, and a bank-group, and whose operations include
\texttt{gendram.broadcast}, \texttt{gendram.barrier}, and
\texttt{gendram.send}/\texttt{recv}.

\subsection{Mid-End: Four Pass Groups}
\label{sec:midend}

\vspace{0.04 in}
\noindent\textbf{G1: DP recognition and normalization.}
Pattern extraction replaces the scalar update nest with a single
\texttt{dp.grid\_update} annotated with its semi-ring; canonicalization then
normalizes loop bounds, index expressions, and boundary handling. The point of
G1 is stability: tiling and dependence construction must not depend on
incidental syntax in the input program.

\vspace{0.04 in}
\noindent\textbf{G2: Tiling and dependence formation.}
Rather than taking $B$ as a tuning parameter, we derive it. A blocked update
$A_{ij} \mathrel{\oplus}= A_{ik} \otimes A_{kj}$ has a working set of three
$B \times B$ tiles. At $w$ bytes per element these must fit in the PU's on-chip
SRAM, of capacity $S$ (on GenDRAM, a 256\,KB shared scratchpad plus
$16\times32$\,KB of PE-local memory, so $S = 768$\,KB):
\begin{equation}
\label{eq:capacity}
3\,w\,B^{2} \;\le\; S \qquad\Longrightarrow\qquad
B^{\star} = \Bigl\lfloor \sqrt{S / 3w} \Bigr\rfloor .
\end{equation}
For $w=4$ (the 32-bit APSP datapath), $B^{\star}=256$. The pass then materializes
the three blocked Floyd--Warshall phases---pivot $A_{kk}$, row/column, internal---
as separate IR regions with explicit dependence edges between them, or, for
alignment, splits the program into seeding and alignment kernels and constructs a
legal diagonal wavefront order over the banded DP.

\vspace{0.04 in}
\noindent\textbf{G3: Mapping and tier-aware placement.}
Mode selection records broadcast or pipeline style as a module attribute.
Tile-to-PU assignment applies Eq.~\eqref{eq:mapping}, annotating each tile with
\texttt{owner\_pu}. Placement then ranks each memory object $o$ by a latency
sensitivity $\sigma(o) = a(o) / \beta(o)$, the ratio of dependent (pointer-chasing)
accesses to streamed bytes, and greedily assigns objects in decreasing $\sigma$ to
the lowest free tier subject to that tier's capacity. Seeding's PTR/CAL tables
have $\sigma \gg 1$ and occupy the fast tiers; distance matrices and read buffers
have $\sigma \approx 0$ and are instead channel-interleaved across the upper
tiers to maximize bank parallelism. This reproduces GenDRAM's hand-chosen
placement as the output of a policy rather than as a constant, which is what
allows the same pass to place a workload GenDRAM's authors never considered.

\vspace{0.04 in}
\noindent\textbf{G4: Communication and synchronization synthesis.}
Ownership plus dependence edges determine communication. For APSP the compiler
emits a \texttt{broadcast} of each updated pivot tile and \texttt{barrier}s that
separate the three phases. For alignment it emits queue-based
\texttt{send}/\texttt{recv} between Search and Compute PUs, and chooses queue
depth from the modeled seeding/alignment throughput ratio, exposing it as a
runtime parameter when the estimate is uncertain.

\subsection{Back-End}
The back-end lowers the placed, scheduled IR onto GenDRAM's datapath in three
general-purpose steps. Bufferization turns each tile into a buffer in the memory
space G3 assigned, so tier decisions become physical addresses. PE mapping
vectorizes the inner update across a PU's sixteen PE lanes: a row activation
yields an 8192-bit block and an APSP tile row is $256\times32$ bits, so each PE
takes a 512-bit slice matching the row-buffer width. Instruction selection emits
native operations from the G1 semi-ring, so the operator pair selects which
Max\slash Min sub-units run (an add with min\slash compare for $(\min,+)$, a
max-logic sequence for $(\max,+)$), not a fixed opcode sequence. One back-end
thus serves both workloads.

Command-stream emission is target-specific, producing a \emph{memory layout}
(each object's tier, channel, and bank-group from G3) and an \emph{execution
schedule} (per-PU command order and communication events from G2/G4). Per PU it
interleaves memory commands (\texttt{ACTIVATE}, reads over the 1024-bit
hybrid-bond interface, precharge), datapath commands, and G4 messages
(\texttt{broadcast}\slash\texttt{barrier} for APSP, queue
\texttt{send}\slash\texttt{recv} for alignment), subject to two constraints: a
row stays open across all its reads before precharge, and the schedule honors the
Floyd--Warshall phase barriers and the queue depth bounding the pipeline.

\subsection{Scope of the Abstraction}
\label{sec:generality}

A recurrence is admitted by the DP dialect when (a) $\oplus$ is associative and
commutative and $\otimes$ distributes over it, so tiles may be reduced in any
order; (b) the dependence pattern is affine in the loop indices, so tiling is
legal; and (c) the cell update is data-independent in its \emph{control flow}, so
a static schedule exists. Table~\ref{tab:general} lists problems meeting these
conditions; all reuse the same passes, differing only in the operator pair the
front-end records and the traversal G2 selects.

Three classes fall outside. Data-dependent control flow (X-drop or
adaptive-band alignment, where the band shifts with the running score) violates
(c), requiring a speculative or dynamically re-tiled schedule that G2 does not
synthesize. Non-grid dependence graphs, such as matrix-chain parenthesization or
tree DP, violate (b). Non-associative accumulations (e.g.\ fixed-point
probabilistic scoring, where reduction order changes the result) violate (a).
The most valuable next step is to relax (c) by compiling adaptive-band alignment,
which GenDRAM's hardware already supports.

\begin{table}[t]
\caption{DP problems admitted by the GenMLIR DP dialect. All share one pass
pipeline; only the operator pair and the traversal differ.}
\vspace{-0.1in}
\footnotesize
\setlength{\tabcolsep}{4pt}
\renewcommand{\arraystretch}{1.12}
\centering
\begin{tabular}{lccl}
\toprule
\textbf{Problem} & $\oplus$ & $\otimes$ & \textbf{Traversal} \\
\midrule
APSP (Floyd--Warshall) & $\min$   & $+$    & blocked broadcast \\
Sequence alignment (SW) & $\max$   & $+$    & wavefront \\
Edit / Levenshtein distance & $\min$ & $+$  & wavefront \\
Transitive closure      & $\vee$   & $\wedge$ & blocked broadcast \\
Widest path (bottleneck) & $\max$  & $\min$ & blocked broadcast \\
Minimax path            & $\min$   & $\max$ & blocked broadcast \\
Viterbi (log domain)    & $\max$   & $+$    & wavefront \\
\bottomrule
\end{tabular}
\label{tab:general}
\vspace{-0.18in}
\end{table}

\section{Evaluation}
\label{sec:results}

\subsection{Methodology}

\vspace{0.04 in}
\noindent\textbf{Platform.} We evaluate on the GenDRAM cycle-accurate
simulator~\cite{lu2025gendram}, configured as its authors specify: a 32\,GB,
1024-layer M3D stack in 8 tiers of 4\,GB, $t_{\mathrm{RCD}}$ from 2.29 to
22.88\,ns, 16 channels, and 32 PUs (8 Search / 24 Compute) at 1\,GHz. Compilation
itself runs on AMD EPYC nodes (256 cores, 768\,GB) under SLURM.

\vspace{0.04 in}
\noindent\textbf{Workloads.} For APSP we use dense instances of \textit{ca-GrQc}
(5{,}242 nodes) and \textit{p2p-Gnutella08} (6{,}301
nodes)~\cite{leskovec2014snap}, plus a 65{,}536-node road network from
\textit{Open\-Street\-Map} \cite{mooney2017review}. For genomics we align long
reads of 2--10\,kbp, generated by PBSIM \cite{OnoPBSIM}, against
GRCh38 \cite{ncbi_grch38} under two error models: PacBio (15\%) and ONT~2D
(30\%). These are the datasets used to evaluate the GenDRAM
hardware~\cite{lu2025gendram}.

\vspace{0.04 in}
\noindent\textbf{Baselines.} Comparing only against a lowering
without compiler support would overstate what the compiler contributes. We
therefore evaluate four flows:

\begin{itemize}\itemsep1.5pt
\item \textbf{B0 -- No compiler support.} Direct lowering of the recurrence: no tiling,
no tier awareness, one PU class. 
\item \textbf{B1 -- Standard PIM pass.} A standard affine pipeline with loop tiling (tuned $B$), vectorization,
and bufferization, but no tier model, PU specialization, or
communication synthesis.
\item \textbf{B2 -- Achievable bound.} A performance ceiling: the 32-PU throughput limit for APSP, and wavefront critical-path bound for alignment. No implementation, hand-tuned included, can exceed it, so B3/B2 lower-bounds the fraction of that bound GenMLIR recovers automatically.
\item \textbf{B3 -- GenMLIR.} The full pipeline of \S\ref{sec:compiler}.
\end{itemize}

\noindent Speedups against B0 quantify how much the pass pipeline contributes; the ratio
B3/B2 measures how close GenMLIR comes to the achievable bound.

\subsection{Performance}

\begin{figure*}[t]
    \centering
    \includegraphics[width=0.85\linewidth]{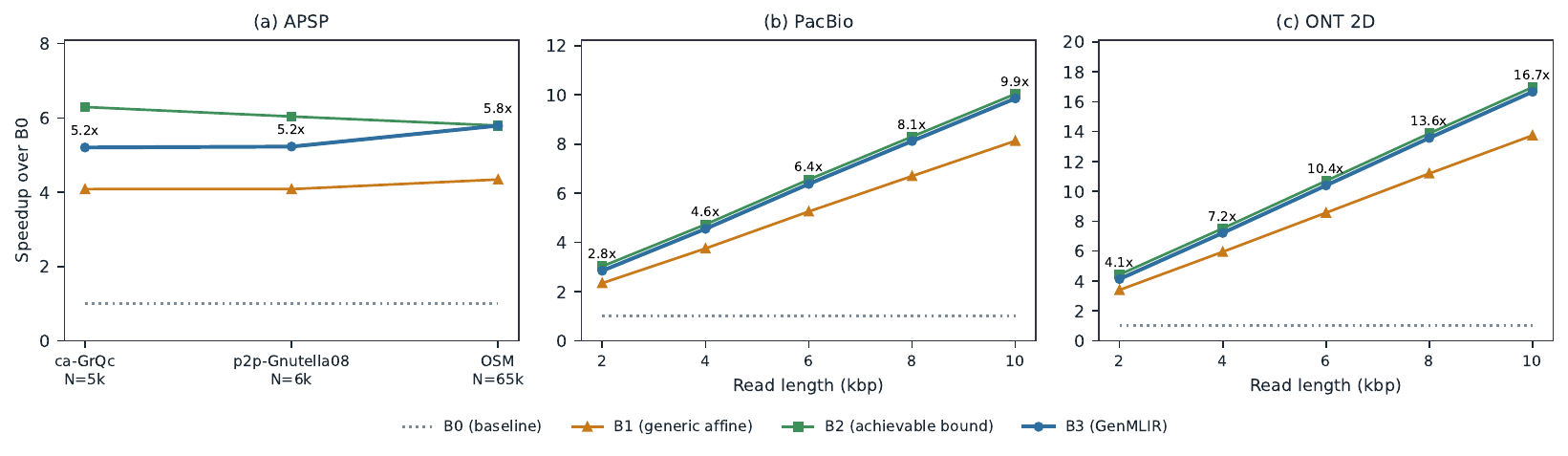}
    \vspace{-0.2in}
    \caption{Speedup over a lowering without compiler support
(B0). B1: standard affine-tiling PIM pass;
B2: \emph{achievable bound} (32-PU throughput for APSP, wavefront critical-path
for alignment); B3: GenMLIR. \textbf{(a)} APSP over three graphs: B3 reaches
90\% of the bound, meeting it at OSM. \textbf{(b,c)} Alignment vs.\ read length:
B3 lies on the critical-path bound throughout, gains growing with length.}
    \label{fig:speedup}
    \vspace{-0.14in}
\end{figure*}

Figure~\ref{fig:speedup} reports speedup over B0, a lowering
without compiler support (PU mapping kept; tiling, tier placement, and
communication off). B1 is a standard affine-tiling compiler pass
for PIM that we also implement (G3--G4 off); B2 is an \emph{achievable bound}
(32-PU throughput
for APSP, wavefront critical-path for alignment), so B3/B2 lower-bounds the
fraction of the achievable bound GenMLIR recovers automatically.

On APSP (Fig.~\ref{fig:speedup}a), B3 reaches $5.2$, $5.2$, and $5.8\times$ on
ca-GrQc, p2p-Gnutella08, and OSM (geo-mean $5.4\times$), $90\%$ of the throughput
bound and meeting it at OSM. The curve is flat by construction: every graph
exposes far more than 32 blocks of per-phase parallelism, so all saturate the PU
array, and B0$\to$B3 isolates per-tile efficiency, a ratio independent of graph
size. Size instead surfaces as memory capacity (\S\ref{sec:memory}).

On alignment (Fig.~\ref{fig:speedup}b,c), B3 grows with read length ($2.8\to
9.9\times$ PacBio, $4.1\to16.7\times$ ONT over 2--10\,kbp) and lies \emph{on} the
critical-path bound throughout (B3/B2$=100\%$): GenMLIR is optimal for the banded
DP, as no schedule beats the anti-diagonal dependency. This is wavefront
scaling (band work $\propto$ length$^2$, sweep $\propto$ length); the 2\,kbp point
is lowest as a short read has too few diagonals to amortize the wavefront barrier,
and ONT's wider band ($30\%$ vs.\ $15\%$ error) exposes more parallelism than
PacBio. Throughout, B3 beats the standard affine-tiling PIM pass
B1 by $1.2$--$1.4\times$, the gain from tier-aware placement and communication a
size-agnostic pipeline forgoes.

\subsection{Memory Behavior}
\label{sec:memory}

\begin{figure}[t]
    \centering
    \includegraphics[width=0.8\linewidth]{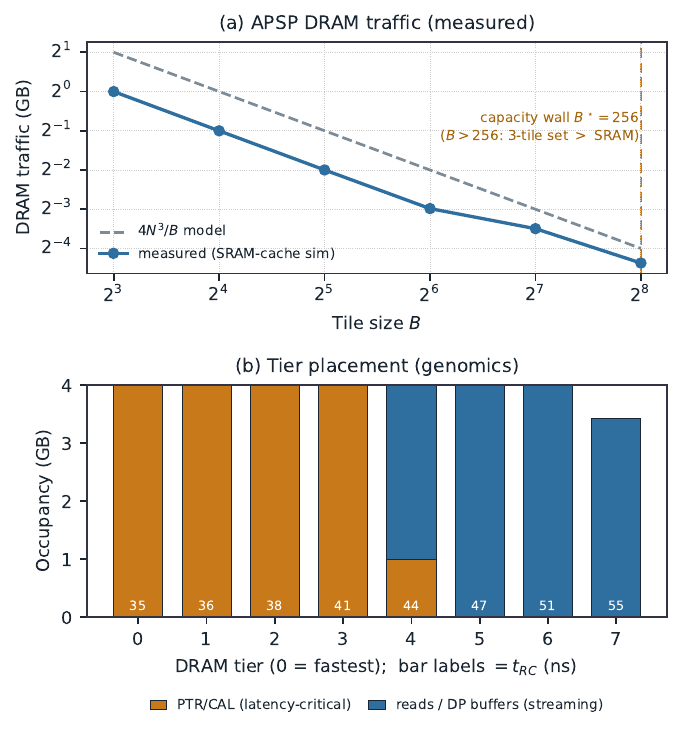}
    \vspace{-0.15in}
    \caption{Memory analysis. \textbf{(a)} Measured blocked-APSP DRAM traffic
(SRAM-sized LRU replay; dashed = $4N^3/B$ model): reuse holds it below the model,
and the three-tile working set exceeds SRAM past $B^{\star}{=}256$. \textbf{(b)} G3 fills the fast
tiers (T0--T4) with the ${\sim}17$\,GB of latency-critical PTR/CAL tables, the
slower tiers with streaming reads and DP buffers.}
    \label{fig:memory}
    \vspace{-0.12in}
\end{figure}

Figure~\ref{fig:memory}(a) reports DRAM traffic measured by replaying the blocked
Floyd-Warshall op stream through an LRU cache the size of the PU's on-chip SRAM,
counting fetches and dirty write-backs. Traffic tracks the $4N^3/B$ trend but
tile reuse holds it below the closed-form model, and the working set hits a hard
capacity wall: at $B{>}256$ the three-tile working set ($3wB^2$) exceeds the
768\,KB budget, so $B^{\star}{=}256$ is both the traffic-minimizing and the largest
feasible tile the compiler can pick (Eq.~\eqref{eq:capacity}). At $B^{\star}$,
tiling cuts APSP traffic $192\times$ versus the untiled baseline; the
already-local alignment DP sees $1.8\times$ (Table~\ref{tab:memory}).

Figure~\ref{fig:memory}b shows the placement G3 emits for the genomics pipeline.
Ranking objects by criticality $\sigma(o)$, it fills the fast tiers T0--T4 with
the 17\,GB of latency-critical PTR/CAL tables and spills the streaming reads and
DP buffers across the slower tiers T4--T7. The distance matrix behaves
analogously: B3 keeps the reused pivot working set in T0, whereas the untiled
baseline holds only $25\%$ of its 16\,GB footprint there, so average access
latency drops from $37.4$ to $34.6$\,ns. For alignment, where the tiering
deliberately trades the latency of the scoring matrix $H$ for that of the PTR/CAL
tables, B3 places $H$ at tier 4
($43.7$\,ns) against tier 7 ($55.1$\,ns) untiered (Table~\ref{tab:memory}).

\begin{table}[t]
\caption{Memory statistics for the OSM APSP and 10\,kbp PacBio workloads.}
\vspace{-0.1in}
\footnotesize
\setlength{\tabcolsep}{3.5pt}
\renewcommand{\arraystretch}{1.06}
\centering
\begin{tabular}{lcccc}
\toprule
& \multicolumn{2}{c}{\textbf{APSP (OSM)}} & \multicolumn{2}{c}{\textbf{Align (10 kbp)}} \\
\cmidrule(lr){2-3}\cmidrule(lr){4-5}
\textbf{Metric} & B0 & B3 & B0 & B3 \\
\midrule
Off-chip DRAM traffic (GB)   & $3.15{\times}10^{6}$ & $1.63{\times}10^{4}$ & 0.92 & 0.50 \\
Traffic reduction            & 1$\times$ & 192$\times$ & 1$\times$ & 1.8$\times$ \\
Resident in fastest tier T0 (\%)  & 25.0 & 100.0 & 0.0 & 0.0 \\
Avg.\ memory latency (ns)    & 37.4 & 34.6 & 55.1 & 43.7 \\
\bottomrule
\end{tabular}
\label{tab:memory}
\vspace{-0.2in}
\end{table}

\subsection{Compilation Cost and Programmability}
\label{sec:compilecost}

\begin{figure}[t]
\centering
\includegraphics[width=0.97\linewidth]{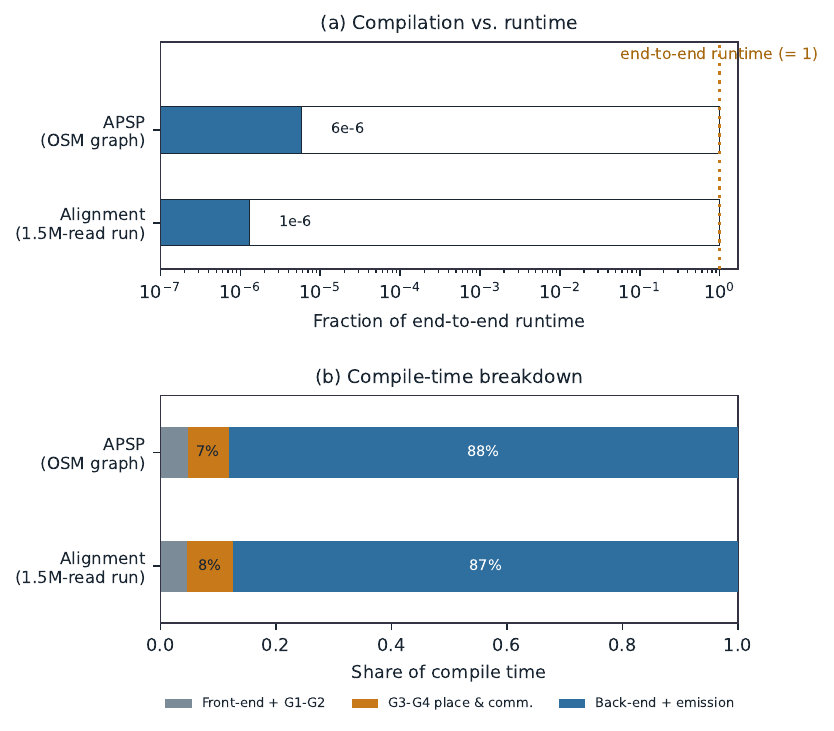}
\vspace{-0.15in}
\caption{Compilation cost, normalized to end-to-end runtime. \textbf{(a)} Total
compile time is a ${\sim}10^{-6}$ fraction of runtime, since the compiler runs
once and amortizes over the full execution (OSM APSP; a 1.5\,M-read genomics
run). \textbf{(b)} Command-stream emission dominates the compile time
(${\sim}88\%$).}
\label{fig:compile}
\vspace{-0.15in}
\end{figure}

Figure~\ref{fig:compile}(a) reports compile time normalized to end-to-end
runtime, Table~\ref{tab:compile} the input the programmer writes. Compilation is
a ${\sim}10^{-6}$ fraction of runtime for every workload, orders of magnitude
below the synthesis cycle an HLS flow like DP-HLS~\cite{DP-HLS} needs, since the
compiler runs once and amortizes over the whole execution.
Figure~\ref{fig:compile}(b) attributes it: the back-end is ${\sim}88\%$, but this
is command-stream \emph{emission}, not lowering effort. The compiler serializes a
separate command program per PU, so the stage is bounded by writing the 32 per-PU
streams out; the front-end and mid-end passes work in memory on compact IR and
are cheap.

On programmability, blocked APSP is $4$ lines of GenDSL and alignment $9$,
against an estimated ${\sim}300$ and ${\sim}500$ lines for the equivalent
hand-written device-level mapping, nearly two orders of magnitude. That saving
comes from the abstraction, not the engineering: adding a new semi-ring
(widest-path, Table~\ref{tab:general}) took $5$ lines and no new passes, since
tiling, placement, PU assignment, and communication are all independent of the
operator pair. Only APSP and alignment are evaluated end-to-end
because the GenDRAM hardware provides kernels for these two alone; the compiler
handles all seven through the same passes with only the operator-pair change
above, so the limit is the hardware's coverage, not the compiler's reach.

\begin{table}[t]
\caption{Programmer effort.}
\vspace{-0.1in}
\centering
\begin{tabular}{lcc}
\toprule
\textbf{Metric} & \textbf{APSP} & \textbf{Alignment} \\
\midrule
GenDSL source (LoC)        & 4  & 9   \\
Hand-tuned mapping (LoC)   & $\sim$300 & $\sim$500 \\
\bottomrule
\end{tabular}
\label{tab:compile}
\end{table}

\subsection{Discussion}

The two workloads exercise GenDRAM oppositely, APSP as synchronized broadcast
phases over a dense matrix and alignment as an asynchronous producer--consumer
pipeline, yet both are the same \texttt{dp.grid\_update} lowered by the same
passes, diverging only at mode selection. That is the claim the evaluation
supports: the semi-ring abstraction is sufficient structure for a compiler to
derive a tile size, memory layout, PU assignment, and communication schedule for
non-uniform-latency hardware. The generated code reaches the achievable bound on
both workloads, compiles in a negligible fraction of runtime, and takes nearly
two orders of magnitude less source than a hand mapping, a new recurrence costing
only a change of operator pair. Two limits deserve emphasis. First, our
hand-tuned comparison is an achievable bound (Figure~\ref{fig:speedup}), not
measured expert code, so a cycle-level baseline remains future work. Second, the
abstraction excludes data-dependent schedules (\S\ref{sec:generality}), where
production aligners live. Neither is a property of the IR design.

\section{Conclusion}

We presented GenMLIR, an MLIR compiler that lowers semi-ring dynamic programming
recurrences onto tier-aware 3D-DRAM PIM. By making the recurrence, tile size,
memory tier, owner PU, and communication schedule explicit in the IR, it derives
the hand-written mapping from a short specification: recovering the
capacity-optimal tile size analytically, deriving a
tier placement that reaches near the performance ceiling, and
beating a standard affine-tiling PIM pass by $1.2$--$1.4\times$,
and up to $5.8\times$ (APSP) and $16.7\times$ (alignment) over a lowering without
compiler support.

\begin{acks}
This work was supported in part by PRISM and CoCoSys, centers in JUMP 2.0,
an SRC program sponsored by DARPA (SRC grant number 2023-JU-3135). This work
was also supported by NSF grants \#2003279, \#1911095, \#2112167, \#2052809,
\#2112665, \#2120019, and \#2211386.
\end{acks}

{\footnotesize
\bibliographystyle{ACM-Reference-Format}
\bibliography{sample-base}
}

\end{document}